\documentclass[11pt,a4paper]{article} 
\usepackage{amssymb,amsmath}
\usepackage{amsthm}
\usepackage{comment}
\usepackage{url}

\usepackage{bm}
\usepackage{MnSymbol, soul}

\newcommand{\ave}[1]{ {\rm E}_p \left[ #1 \right]}

\newcommand{\vtheta}{ \boldsymbol{\theta} }
\newcommand{\veta}{ \boldsymbol{ \eta} }

\title{A short note on basic of information geometry from thermodynamics and thermostatistics}

\author{Tatsuaki Wada}

\date{sul $1^{\mathrm o}$ settembre 2021}

\begin{document}
\maketitle

\begin{abstract}
This is a short note for some basics of information geometry from thermodynamics and Callen's themostatistics.
\end{abstract}

\section{Introduction}

  The geometric approaches to thermodynamics and statistical mechanics have been
  developed since the early works of Gibbs \cite{Gibbs} and Carath\'eodory \cite{ Cara}.
  Ruppeiner \cite{Ruppeiner} and Weinhold \cite{Weinhold} independently introduced the Riemannian metrics which are
  constructed from thermodynamic potentials (entropy or internal energy).
  The thermodynamic fluctuations around the equilibrium states
  have been studied,
  and the associated Riemannian curvature has been related to an interaction which
  characterizes a thermodynamic system.
  On the other hand, information geometry \cite{AN00} has been developed
  mainly in the fields of statistics, and it provides a useful framework for
  studying the family of probability distributions, e.g., the so-called \textit{exponential family}, by using the geometric tools in affine differential geometry.
One of the distinct features in information geometry is a dualistic structure of affine connections, which provides us a very useful tool
for many scientific fields, such as information theory, statistics, neural networks, statistical physics, and so on.

In the next section we begin with a brief review of the geometric approach to
thermodynamics and Callen's thermostatistics \cite{Callen}.
Section 3 provides the preliminaries on the Hessian geometry concerning the information geometry based on the exponential family.

\section{Thermodynamics and thermostatistics}

In equilibrium thermodynamics, a thermal system is described by a small number of extensive variables such as the entropy $S$,
the internal energy $U$, the volume $V$, the total number of particles $N$ and so on.
These extensive variables are not independent, but linked by a so-called fundamental equation which customarily written 
in the energy representation as
\begin{align}
  U = U(S, V,  N, \ldots ),
\end{align}
or in entropy representation 
\begin{align}
  S = S(U, V,  N, \ldots).
 \end{align}
 
Here we use the entropy representation, and for the sake of simplification we consider the case
in which the entropy is a function of only the internal energy $U$ and the volume $V$, i.e., $S=S(U, V)$.
As well known, the first law of thermodynamics is expressed as
\begin{align}
  dS(U, V) = \frac{1}{T} \, dU + \frac{P}{T} \, dV,
\end{align}
where $T$ and $P$ denote the temperature and the pressure of the thermal
system, respectively. They are intensive variables and related by the following relations:
\begin{align}
  \frac{1}{T} = \left( \frac{\partial S}{\partial U} \right)_{V},\quad
  \frac{P}{T} = \left( \frac{\partial S}{\partial V} \right)_{U}.
  \label{thermo-rel}
\end{align}

Mathematically these relations are necessary and sufficient conditions
so that the Pfaffian equation 
\begin{align}
 \omega := dS(U,V) -\frac{1}{T} \, dU - \frac{P}{T} \, dV =0,
 \label{omega}
 \end{align}
is \textit{completely integrable}, and
consequently the entropy $S(U,V)$ is a state function as shown originally
by Carath\'eodory \cite{Cara}.
Here $\omega$ is a differential one-form and, according to Frobenius' theorem, a necessary and sufficient condition \cite{Pfaffian} for the Pfaffian equation $\omega = 0$ is
\begin{align}
  d \omega \wedge \omega = 0,
  \label{Frobenius}
\end{align}
in general.
For the Pfaffian equation \eqref{omega}, the condition \eqref{Frobenius} becomes
\begin{align}
d \omega \wedge \omega&=
  \left\{ -d \left( \frac{1}{T} \right) \wedge dU - d\left( \frac{P}{T} \right) \wedge dV \right\} \wedge
  \left\{  \left( \frac{\partial S}{\partial U} \right)_{V} \, d U +  \left( \frac{\partial S}{\partial V} \right)_{U} \,dV -\frac{1}{T} \, dU - \frac{P}{T} \, dV \right\} 
  \notag \\
  &= \left\{ \frac{1}{T} - \left( \frac{\partial S}{\partial U} \right)_{V} \right\} d\left( \frac{P}{T} \right) \wedge dV \wedge dU
     + \left\{ \frac{P}{T} - \left( \frac{\partial S}{\partial V} \right)_{U} \right\} d\left( \frac{1}{T} \right) \wedge dU \wedge dV = 0.
  \end{align}
Consequently the relations \eqref{thermo-rel} are obtained.

Next we introduce the
Planck potential $\Xi$ given by
\begin{align}
  \Xi\left( \frac{1}{T}, \frac{P}{T} \right)
  = S(U, V) - \frac{1}{T} \, U - \frac{P}{T} \, V,
  \label{Planck}
\end{align}
which is nothing but the total Legendre transform of $S(U,V)$.
For the sake of later convenience, instead of the concave function $S(U, V)$,
we use the convex function $-S(U, V)$, which is called \textit{negentropy}, or negative entropy \cite{Brillouin}.
Introducing the set of the extensive variables $\eta_i, (i=1,2)$
with  $\eta_1 = U, \eta_2 = V$,
and the set of the intensive variables $\theta^i, (i=1,2)$
with $\theta^1 = -1/T, \theta^2 = -P/T$,
Relations \eqref{thermo-rel} can be compactly expressed as
\begin{align}
  \theta^i = \partial^i \big( -S(\eta_1, \eta_2) \big), \quad i=1,2,
  \label{ti}
\end{align}
where $\partial^i = \partial / \partial \eta_i$.
The Legendre relation \eqref{Planck} becomes
\begin{align}
  \Xi(\theta^1, \theta^2) = \sum_{m=1}^2 \theta^m \, \eta_m + S(\eta_1, \eta_2),
  \label{Leg-Planck}
\end{align}
and the dual relation of \eqref{ti} is readily obtained from \eqref{Leg-Planck}
as
\begin{align}
  \eta_i = \partial_i \, \Xi(\theta^1, \theta^2), \quad i=1,2.
  \label{ei}
\end{align}
where $\partial_i = \partial / \partial \theta^i$.

It is also known that the Maxwell relations in thermodynamics are
due to the irrelevance of
the order of differentiating a thermodynamic potential (an analytic
function) with respect to two variables,
For instance, for the negentropy $-S$, the following Maxwell relation
\begin{align}
  \frac{\partial}{\partial V} \, \left(\frac{1}{T} \right)
   = \frac{\partial}{\partial U} \, \left( \frac{P}{T} \right)
\end{align}
is equivalent to the relation:
\begin{align}
  \partial^2 \partial^1 (-S) = \partial^1 \partial^2 (-S).
\end{align}
The Hessian of the negentropy $(-S)$,
can be considered as a symmetric metric tensor of a manifold with 
the thermodynamic variables $\eta_i$ as its coordinates,
\begin{align}
  (g^{\rm R})^{ij} \equiv \partial^i \partial^j (-S),
\end{align}
which is equivalent to the \textit{Ruppeiner metric} \cite{Ruppeiner}.
The inverse matrix of $(g^{\rm R})^{ij}$ is given by
\begin{align}
  g^{\rm R}_{ij} = \partial_i \partial_j \, \Xi.
\end{align}

\subsection{Thermostatistics}

In Callen's thermostatistics \cite{Callen}, the concept of the equilibrium states
in conventional thermal physics is extended to the ``equilibrium states''
which are characterized as the states that maximizes a certain kind of measures of
information.
A well known measure of information is the Gibbs-Shannon entropy $S$, which is expressed as 
the expectation of $-\ln p(x; \vtheta)$, i.e.,
\begin{align}
  S = \ave{-\ln p(x; \vtheta)}.
  \label{GS}
\end{align}
Here and hereafter, $\ave{\cdot}$ stands for the standard expectation with respect
to a pdf $p(x; \vtheta)$,
which is characterized by the parameter $\vtheta = (\theta^1, \theta^2, .....\theta^M)$ with an appropriate degree of freedom $M$.
Any extensive thermal quantity is considered as the expectation (or average)
of the corresponding microscopic quantity, for instance the internal energy is given by
\begin{align}
   U = \int dx \; p(x; T, P) E(x),
\end{align}
where $E(x)$ is the microscopic energy of a configuration $x$, 
and $p(x; T, P)$ is a pdf depending on the intensive parameter $T$ and $P$.
Introducing the notation $f_i(x)$ for the microscopic quantity associated with $\eta_i$,
we can express the extensive thermal quantities as 
\begin{align}
   \eta_i = \int dx \; p(x; \vtheta) f_i(x) = \ave{f_i(x)}, \quad i=1,2, ..., M.
   \label{eta-ave}
\end{align}
From the Legendre transformation \eqref{Leg-Planck} and Equation \eqref{eta-ave}
we see that
\begin{align}
     \ave{ \ln p(x; \vtheta)} = -S(\veta)
  = \sum_m \theta^m \ave{f_m(x)} - \Xi(\vtheta)
 = \ave{ \sum_m \theta^m f_m(x) - \Xi(\vtheta)}.
\end{align}
We thus obtain that \footnote{One can add a fluctuating term $\chi$ with zero expectation  $\ave{\chi} = 0$ to the left hand side. }
\begin{align}
  \ln p(x; \vtheta) = \sum_m \theta^m f_m(x) - \Xi(\vtheta),
  \label{lnp}
\end{align}
and consequently we see that the $p(x; \vtheta)$ is an exponential pdf:
\begin{align}
   p(x; \vtheta) = \exp \left( \sum_m \theta^m f_m(x) - \Xi(\vtheta) \right).
  \label{exp-prob}
\end{align}


The quantity
\begin{align}
  \partial_i \, \ell_{\theta}(x) \equiv \partial_i \, \ln p(x; \vtheta), \quad i=1, 2, \ldots, M,
  \label{score-func}
\end{align}
is called \textit{the score function} in statistics, and it has
zero expectation, i.e.,
\begin{align}
   \ave{\partial_i \, \ell_{\theta}(x)} \equiv 
\int dx \, p(x; \vtheta) \, \partial_i \, \ln p(x; \vtheta)
  = \int dx \, \partial_i p(x; \vtheta) = \partial_i \ave{1} = 0,
 \label{Ex_score}
\end{align}
which is due to the normalization $ \ave{1} = 1$ of any pdf $p(x; \vtheta)$.
From this we readily confirm Relation \eqref{ei} as
\begin{align}
  0 = \ave{\partial_i \ell_{\theta}(x)}
  = \ave{f_i(x)} - \partial_i \, \Xi(\vtheta) = \eta_i  - \partial_i \, \Xi(\vtheta).
  \label{zero}
\end{align}

Let us introduce Fisher's information matrix $g^{\rm F}_{ij}$ defined by
\begin{align}
 g_{ij}^{\rm F} (\vtheta) \equiv \ave{
  \partial_i \ell_{\theta} (x) \, \partial_j \ell_{\theta}(x) }.
 \quad i,j = 1,2, \ldots, M.
 \label{gF}
\end{align}
%
%
Differentiating the both sides of Equation \eqref{Ex_score} with respect to $\theta$,
we obtain
\begin{align}
 \int dx \, \partial_i p(x; \vtheta) \, \partial_j \ell_{\theta}(x) 
= - \int dx \, p(x; \vtheta) \, \partial_i \, \partial_j \ell_{\theta}(x).
\end{align}
Using this relation, the Fisher metric $g^{\rm F}$ can be written equivalently
in other different expressions:
\begin{align}
 g^{\rm F}_{ij} &= \int dx \, \partial_i p(x; \vtheta) \, \partial_j \ell_{\theta}(x) 
\label{rel2}\\
 &= - \int dx \, p(x; \vtheta) \, \partial_i \, \partial_j \ell_{\theta}(x) 
\label{rel3}\\
 &= \int dx \, \frac{1}{p(x; \vtheta)} \, \partial_i p(x; \vtheta) \partial_j p(x; \vtheta).
\label{rel4}
\end{align}
In particular, substituting Equation \eqref{lnp} into \eqref{rel3},
we readily confirm that:
\begin{align}
 g^{\rm F}_{ij} = \partial_i \partial_j \Xi( \vtheta ) = g^{\rm R}_{ij},
\end{align}
that is, this Fisher metric for the exponential pdf \eqref{exp-prob} is a Hessian matrix,
and coincides with the inverse matrix of Ruppeiner metric $(g^{\rm R})^{ij}$.

\section{Basic of information geometry}

In information geometry \cite{AN00}, a pair of dually flat affine connections
play an essential role in the geometrical methods of statistical inference.
A well known dually flat space is the statistical manifold of the exponential family, which can be naturally considered as a Hessian manifold. 

\subsection{Hessian geometry}

We here briefly review the basics of Hessian manifold.
For more details please see Ref. \cite{MH13}.
Let ${\mathcal M}$ be a manifold, $h$ is a positive definite metric, and
$({\mathcal M}, h)$ be a Riemannian manifold.
For an affine connection $\nabla$, we can define the dual connection $\nabla^{\star}$ of $\nabla$
associated with $h$ by 
\begin{align}
  X h(Y, Z) = h(\nabla_{X} Y, Z) + h(Y, \nabla^{\star}_{X} Z),
\end{align}
where $X, Y$ and $Z$ are vector fields on ${\mathcal M}$. The affine connection $\nabla$ is also dual of
$\nabla^{\star}$.
We say that $\nabla$ is curvature-free if the curvature tensor
\begin{align}
  R(X, Y) Z \equiv \nabla_X \nabla_Y Z - \nabla_Y \nabla_X Z - \nabla_{[X, Y ]} Z,
\end{align}
vanishes everywhere on ${\mathcal M}$. Here $[X, Y ] \equiv XY - YX$.
The torsion tensor $\mathcal{T}$ is defined by
\begin{align}
  \mathcal{T}(X, Y) \equiv \nabla_X Y - \nabla_Y X  - [X, Y ],
\end{align}
and we say that $\nabla$ is torsion-free if $\mathcal{T}$ vanishes everywhere on ${\mathcal M}$.

An affine connection $\nabla$ is assumed to be torsion-free in this study.
If an affine connection $\nabla$ is curvature-free, we say the $\nabla$ is flat.
In this case there exists a coordinate system $\{\theta^i \}$ on ${\mathcal M}$ locally
such that the connection coefficients $\{ \Gamma_{ij}^{k} \}$ of $\nabla$ vanish
on the coordinate neighborhood. Such a coordinate system $\{ \theta^i \}$ is
called an \textit{affine coordinate system}.  

For a Riemannian manifold $({\mathcal M}, h)$ and a flat affine connection $\nabla$ on ${\mathcal M}$,
the set $(\nabla, h)$ is called a \textit{Hessian structure} on ${\mathcal M}$ if there exists, at least locally,
a function $\Psi$ such that $h = \nabla d \Psi$.
This is expressed, in the coordinate form, as
\begin{align}
  h_{ij} (\vtheta_p) 
  = \partial_i \partial_j \Psi(\vtheta_p),
\end{align}
where $\vtheta_p$ is the coordinate of an arbitrary point $p$ on ${\mathcal M}$.


It is known that for a Hessian manifold $({\mathcal M}, \nabla, h)$ and the dual
coordinate systems $\{ \theta^i \}$ for $\nabla$ and $\{ \eta_i \}$
for $\nabla^{\star}$, there exists a pair of the potential functions $\Psi$ and $\Psi^{\star}$
on ${\mathcal M}$ such that:

\begin{align}
 \Psi(\vtheta_p) &+ \Psi^{\ast}(\veta_p) - \vtheta_p \cdot \veta_p = 0, 
\label{Legendre0}\\
 \theta^i &= \partial^i \Psi^{\ast}(\veta),\quad
 \quad \eta_i = \partial_i \Psi(\vtheta),\\
 h_{ij} &= \partial_i \partial_j \Psi(\vtheta), \quad
 h^{ij} = \partial^i \partial^j \Psi^{\star}(\veta),
 \label{Legendre}
\end{align}
where the matrix $h_{ij}$ of a Riemannian metric $h(\vtheta)$ is the inverse
matrix $h^{ij}$ of $h(\veta)$, and vise versa.
The potential functions $\Psi(\vtheta)$ and $\Psi^{\ast}(\veta)$ are
Legendre dual to each other, and are called $\theta$- and $\eta$-potential functions, respectively.
Note also that $({\mathcal M}, \nabla^{\star}, h)$ is a Hessian manifold
associated with the potential function $\Psi^{\star}$.

For the exponential pdf:

\begin{align}
   p(x; \vtheta) = \exp \left( \sum_m \theta^m f_m(x) - \Psi(\vtheta) \right),
  \label{exp-pdf}
\end{align}
it is well known that Fisher metric $g^{\rm F}$
is expressed as the Hessian of the potential function, e.g,:

\begin{align}
   g^{\rm F}_{ij}(\vtheta) = \partial_i \partial_j \Psi(\vtheta).
\end{align}
From Equation \eqref{exp-pdf}, we obtain

\begin{align}
  \partial_i \ell_{\theta}(x) = \partial_i \ln p(x; \vtheta) =
  f_i(x) - \partial_i \Psi(\vtheta).
  \label{tangent-vec}
\end{align}
Substituting this relation into Definition \eqref{gF} and
using $\eta_i = \partial_i \Psi(\vtheta)$, we see that

\begin{align}
 g_{ij}^{\rm F} (\vtheta) = \ave{
  \big( f_i - \ave{f_i} \big) \, \big(f_j - \ave{f_j} \big) },
 \quad i,j = 1,2, \ldots, M,
 \label{cov-mat}
\end{align}
which is the covariance matrix characterizing the expectation
of the fluctuations around the expectations $\ave{f_i}$.
Physically this means that the expectations of the relevant thermodynamic
fluctuations characterize the metric $g_{ij}^{\rm F}$ of the statistical
manifold ${\mathcal M}$ for the equilibrium thermodynamics. In addition
since $g_{ij}^{\rm F} = \partial^i \, \eta_j$, each component of $g_{ij}^{\rm F}$
describes a response function which is the derivative of an extensive variable $\eta_j$
with respect to an intensive variable $\theta^i$.
As a result, the physical interpretation of Relation \eqref{cov-mat} is given by 
the fluctuation-response relations
\cite{Callen,WS15} for an equilibrium thermal system. 

The canonical divergence function \cite{AN00} for
the two points $p$ and $r$ on ${\mathcal M}$ can be defined by:

\begin{align}
 D({\boldsymbol p}, {\boldsymbol r})
\equiv \Psi(\vtheta_p) + \Psi^{\star}(\veta_r) - \vtheta_p \cdot \veta_r.
\end{align}
It is well known that for the exponential pdf,
the canonical divergence coincides with the Kullback-Leibler (KL) divergence.
In addition, for the exponential pdf $p(x; \vtheta)$ and an arbitrary pdf $r(x)$, we have

\begin{align}
  \int dx \, r(x) \ln p(x; \vtheta)
  = \int dx \, r(x) \left[ \sum_m \theta^m f_m(x) - \Psi(\vtheta_p) \right]
  = \vtheta_p \cdot \veta_r - \Psi(\vtheta_p).
\end{align}
Then it follows that

\begin{align}
 D({\boldsymbol p}, {\boldsymbol r})
 &= \Psi^{\star}(\veta_r) -\big( \vtheta_p \cdot \veta_r - \Psi(\vtheta_p) \big)
 = \int dx \, r(x) \left[ \ln r(x) - \ln p(x; \vtheta) \right]
\notag \\
& = \int dx \, r(x) \ln \left( \frac{r(x)}{p(x; \vtheta)} \right).
 \label{KL-div}
\end{align}


The dual affine connections $\nabla^{(e)}$ and $\nabla^{(m)}$ are induced
from the Fisher metric.
The Christoffel symbol $\Gamma_{ij, k}^{(e)}$ of the first kind for
the e-affine connection $\nabla^{(e)}$ and that $\Gamma^{(m)}_{ij,k}$ for the m-affine connection $\nabla^{(m)}$ are defined so that the next relation holds
\cite{AN00}:

\begin{align}
 \partial_i g^{\rm F}_{jk} = \Gamma_{ij, k}^{(e)} + \Gamma^{(m)}_{ij,k}.
 \label{i-gF}
\end{align}
More specifically, they are explicitly given by

\begin{align}
 \Gamma_{ij, k}^{(e)} &\equiv 
\int dx \, \partial_k p(x; \vtheta) \partial_i \partial_j \ell_{\theta}(x)
= \ave{ \partial_k \ell_{\theta} \; \partial_i \partial_j \ell_{\theta}(x)},
\label{e-connection}
\\
\Gamma_{ij, k}^{(m)} &\equiv 
 \int dx \, \partial_i \partial_j p(x; \vtheta) \partial_k \ell_{\theta}(x)
= \ave{ \frac{1}{p(x; \vtheta)} \partial_i \partial_j p(x; \vtheta) \; \partial_k \ell_{\theta}(x) },
\label{m-connection}
\end{align}
respectively.

\end{document}